# Structural, dielectric and ferroelectric studies of thermally stable and efficient energy storage ceramic material: $(Na_{0.5-x}K_xBi_{0.5-x}La_x)TiO_3$

Anita Verma[1], Arun Kumar Yadav[1], Nasima Khatun[2], Sunil Kumar[1], Ravindra Jangir[3], Velaga Srihari[4], V. Raghavendra Reddy[5], Shun Wei Liu[6], Sajal Biring[6*], Somaditya Sen[1,2,6 *]

[1] Discipline of Metallurgy Engineering and Materials Science, Indian Institute of Technology Indore, Simrol, Indore 453552, India
[2] Department of Physics, Indian Institute of Technology Indore, Simrol, Indore 453552, India
[3] Synchrotrons utilization section, Raja Ramanna centre for advanced technology, Indore-452013
[4] High Pressure & Synchrotron Radiation Physics Division, Bhabha Atomic Research Centre, 400085, Mumbai, India
[5] UGC-DAE Consortium for Scientific Research, University Campus, Khandwa Road, Indore-452001, India
[6] Electronic Engg., Ming Chi University of Technology, New Taipei City, Taiwan

**Abstract**

The structural, dielectric and ferroelectric properties of lead-free $(Na_{0.5-x}K_xBi_{0.5-x}La_x)TiO_3$ ($0 \leq x \leq 0.12$) powders synthesized by sol-gel self-combustion method were investigated. Rietveld refinement of Synchrotron x-ray diffraction (SXRD) data confirms pure single phase rhombohedral crystal structure with *R3c* space group for all the compositions and anti-phase ($a^-a^-a^-$) octahedral tilting angle decreased with increase in composition *x*. Homogeneity and elemental proportions were confirmed by Energy dispersive x-ray spectrometry (EDS). The temperature-dependent dielectric study has shown two diffuse type of dielectric anomaly for all the samples, due to A-site disorder in the lattice, which has been assigned to two-phase transitions: ferroelectric to anti-ferroelectric and anti-ferroelectric to the paraelectric phase transition. The transition temperature of these phase transitions is found to decrease as a function of composition (*x*). Thermal stability range ($\Delta T$) of dielectric constant increases from ~100 °C to 220 °C as a function of composition (*x*). Stable dielectric constant first increases, from 1557±10 % for parent compound, with the composition, highest for *x* = 0.06 composition with $\varepsilon_{mid}$ ~ 2508±10 % for the temperature range ~180 °C to 340 °C, and after that decreases to 1608±10 % for *x* = 0.12, but remain higher than the parent compound $Na_{0.5}Bi_{0.5}TiO_3$ (NBT). Ferroelectric measurements have shown monotonously decreasing coercive field ($E_c$) as a function of the composition due to decrease in grain size, confirmed by microstructural studies using Field Emission Scanning Electron Microscope (FESEM). Exponential increases in the energy storage efficiency ($\eta$) from ~ 17 % to 87 % as a function of composition (*x*) have also observed. These types of materials, with stable high dielectric constant ($\varepsilon_r$) and low loss ($\tan\delta$), have a vast scope in the field of the thermally stable dielectric constant materials and energy storage applications.

**Introduction**

In recent times, lead-free dielectric materials are gained much attention, due to the toxicity of lead-based materials, with thermally stable dielectric constant (>200 °C) and low dielectric loss. These materials are extensively used as a capacitor where heat dissipation is a major issue mainly in electronic devices used in a harsh environment. Ceramic capacitors with fast charging and discharging rate, better mechanical and thermal properties, which are competitive to battery and other power storage devices [1], though, energy storage density of the ceramic capacitor is less in comparison to the battery. This new type of capacitor materials is used in sensing and controlling system of aerospace, aviation and automotive sectors in where controlling and sensing systems are placed closed to the engine at elevated temperature (~200-300 °C) [2]. These types of capacitor are also useful in down drilling in oil and gas industry in controlling and monitoring, where high-temperature stable electronics are required. Commercially available glass, polymer, and $BaTiO_3$ based materials are useful only below 200 °C [3, 4], whereas, lead-based dielectric materials are used in 200~250 °C temperature range. However, their toxicity and harmfulness to the environment necessitated to look for alternatives.

Sodium bismuth titanate, $Na_{0.5}Bi_{0.5}TiO_3$(NBT), ceramics are one of the most promising lead-free alternatives, with $ABO_3$ type structure, dielectric and ferroelectric materials, discovered in 1960 [5, 6]. It has been used in many devices like sensor, actuators, pyroelectric infrared sensors, etc. [7-11]. NBT crystallizes with a perovskite-like structure with a sequence of phase transitions, a high-temperature NBT is in cubic (space group $Pm\bar{3}m$) structure which transforms to tetragonal (space group $P4bm$) (at 540 °C) and to room-temperature ferroelectric rhombohedral (space group $R3c$) structures (at 320 °C). NBT has A-site disordered structure due to heterovalent atoms at A-site $Na^+/Bi^{3+}$, where A-site equally shares by ferroelectric $Bi^{3+}$(50%) and inactive $Na^+$ (50%) [12, 13]. $Bi^{3+}$ ions have "stereochemically active lone pair electrons", a similar electronic configuration like $Pb^{2+}$ [14], due to which $Bi^{3+}$ has second-highest polarizability after $Pb^{2+}$ [15]. At room temperature NBT has structural distortions due to hybridization between $Bi(6s^2)$ lone pairs and $O(2p)$ electrons, this distortion promotes large ferroelectricity and piezoelectricity in NBT [16-20]. This material has been explored during the past four decades due to its excellent dielectric and ferroelectric properties. It has relatively large remnant polarization, high Curie temperature 320 °C and it also shows two diffuse type of dielectric anomaly with temperature, which is related to two-phase transition ferroelectric (FE) to anti-ferroelectric (AFE), and anti-ferroelectric (AFE) to paraelectric (PE). These phase transition temperatures become more diffuse by hetero-valent substitution at A or B-site in $ABO_3$ lattice system; this is claimed to be a possible reason behind stable dielectric constant in NBT [21, 22]. In this type of dielectric materials, temperature-dependent phase transitions (FE - AFE and AFE - PE) takes place in different micro-regions at different temperatures leads to diffusion of the dielectric peak [23]. These materials are used in the fields where thermally stable dielectric materials required. However, NBT has some drawbacks like high coercive field and leakage current, because of which poling is difficult. Many researchers tried A or B- site substitutions in NBT ceramic to improve its functionality [24]. Several reports are available in literature for modified NBT material with thermally stable dielectric constant, $0.9(Na_{0.5}Bi_{0.5})TiO_3-0.1KTaO_3$ shows a stable dielectric constant ~1250 in wide temperature range -55 °C to 300 °C [25], *Zr* modified $Na_{0.5}Bi_{0.5}TiO_3-Ba_{0.8}Ca_{0.2}TiO_3-NaNbO_3$ shows a stable dielectric constant ~1300 in wide temperature range 80 °C to 280 °C [26], Ca modified $Na_{0.5}Bi_{0.5}TiO_3-BaTiO_3$ shows a stable dielectric constant ~1300 in wide temperature range -50 °C to 200 °C [21].

From above discussion, it is evident to note that substitution at A or B-sites in NBT by different anions create disorder in the lattice, which leads to various interesting phenomena. Electronic polarizability ($\alpha_D$) of *La* ($\alpha_D$ ~ 170 – 214) and *K* ($\alpha_D$ ~ 289 – 294) is larger than *Bi* ($\alpha_D$ ~ 50-60) and *Na* ($\alpha_D$ ~ 161 – 163) respectively [27], apart from this, average crystal radius of $K^+$(1.78Å) and $La^{3+}$(1.5Å) is almost comparable to $Na^+$(1.53Å) and $Bi^{3+}$(1.52Å), there is a probability of A-site substitution in parent compound NBT. This may lead to increase in the dielectric constant of the NBT compound. It is also reported that doping of rare-earth metal $La^{3+}$ suppress the grain growth which reduces coercive field ($E_c$) [28]. Thus, the motivation of this work is to explore the effect of substitution of $K^+/La^{3+}$ on A–site in place of *Na/Bi* on structural, morphology, dielectric and ferroelectric properties. For this, $(Na_{0.5-x}K_xBi_{0.5-x}La_x)TiO_3, (0 \leq x \leq 0.12)$ were synthesized by sol-gel method. Thermally stable dielectric constant and also high energy storage efficiency 87% was observed for $x = 0.12$ composition. This material system can be useful in high-temperature stable capacitors beyond operating temperature >200 °C and efficient energy storage applications.

**Experimental**

A series of polycrystalline $(Na_{0.5-x}K_xBi_{0.5-x}La_x)TiO_3$, ($x = 0, 0.03, 0.06, 0.09,$ and $0.12$) [abbreviated as NKBLT-*x*] powders were prepared using sol-gel self-combustion method. All the precursors used in the synthesis are purchased from Alfa Aesar and used without further purification. Precursors selected to synthesize NKBLT-*x* were sodium nitrate (purity 99.9%), bismuth nitrate (purity 99.9%), dihydroxy-bis (ammonium lactate) titanium (IV) (50% w/w aqua solution (purity 99.9%)), potassium nitrate (purity 99.9%), and lanthanum nitrate (purity 99.9%). These precursors were selected due to their solubility in deionized (DI) water. Bismuth nitrate is not soluble in DI water but is soluble in dilute $HNO_3$. Stoichiometric solutions of each precursor were prepared with DI water in separate beakers. Titanium solutions were mixed in lanthanum solutions, followed by bismuth, sodium, and potassium. These mixed solutions (precursors) were stirred for 2 h on a magnetic stirrer at room temperature. Solutions of ethylene glycol and citric acid of 1:1 molar ratio were prepared in separate beakers as fuel. The fuel solution was added to precursor solution and stirred vigorously, the ratio of oxidizer to the fuel was calculated using the standard procedure mentioned elsewhere [29]. The resultant solution was continuously stirred and heated to form the gel at ~85 °C. The gel was burnt in a fume hood on the hot plate to form dark brown powders. These powders were further heated at 450 °C for 12 h for denitrification and de-carbonization. The resultant powders were calcined at 700 °C for 10 h after grinding in mortar and pestle. These powders were mixed with 5 weight % polyvinyl alcohol solution (binder) and pelletized with a uni-axial press into discs of ~13 mm diameter and ~1.5 mm thickness. The binder was burnt off at 600 °C for 6 h. Further sintering continued at 1150 °C for 3 h to form dense pellets.

Synchrotron powder x-ray diffraction (SRPXRD) measurements were carried out on the ensuing powders, at ambient conditions, at Extreme Conditions Angle Dispersive/Energy Dispersive X-Ray Diffraction (EC-AD/ED-XRD) Beam Line-11 (BL-11) at Indus-2 synchrotron source, RRCAT, Indore, India. Measurements were performed in rotating capillary mode at ~150 rpm to reduce orientation effects. Diameters of capillary tubes were selected taking into account the linear absorption coefficient of samples and 25% packing density of material in a capillary. The desired wavelength, λ = 0.50346 Å, for ADXRD diffraction experiments was selected from the white light of the bending magnet using a Si(111) channel-cut monochromator. The monochromatic beam is then focused on to the samples with a Kirkpatrick-Baez (K-B) mirror. A MAR345 image plate detector (which is an area detector) was used to collect two-dimensional diffraction data. The sample to detector distance and wavelength of the beam were calibrated using NIST standards $CeO_2$ and $LaB_6$. Calibration and conversion/integration of 2D diffraction data to 1D, intensity versus 2θ, was carried out using FIT2D software [30]. Microstructure and elemental analysis of sintered pellets were investigated by Supra55 Zeiss field emission scanning electron microscope equipped with energy dispersive x-ray spectrometer (EDS). Experimental bulk densities, $ρ_e$, of NKBLT-$x$ pellets were estimated by Archimedes' methods using Xylene (density=0.86 g/cm$^3$) as the liquid media. Electrodes were prepared using high-temperature silver paste for electrical property measurements. Silver paste was painted on both sides of sintered pellets. These pellets were cured at 550 °C for 15 min. To avoid moisture content in the samples finally, the pellets were annealed at 200 °C for 15 min before electrical measurements. The dielectric response was measured using a Newtons 4$^{th}$ LTD phase sensitive multimeter with the signal strength of 1V$_{rms}$. A ferroelectric loop (P-E) tracer of M/s Radiant Instruments, USA was used to perform ferroelectric (P-E) loop measurements on sintered pellets immersed in silicone oil (to prevent electric arcing at high voltages).

**Results and Discussion**

Measured SRPXRD patterns of *NKBLT-x* powders are shown in Fig.1. Preliminary analysis of the SRPXRD patterns shows that all the samples are in rhombohedral structure with R3c space group as reported in the case of parent NBT. Insets in Fig. 1 shows enlarged the portion of the SRPXRD pattern for Bragg peaks (110) and (113), though similar changes are observed with other Bragg reflections as well, we showed these Bragg peaks as a representation. It is seen from the Bragg peak (110), that the diffraction peak shifts to lower 2theta as the value of $x$ increases, which indicated lattice expansion in *ab*-plain. Observation of (113) Bragg peak (inset in Fig. 1) indicated that with the increase in substitution, the intensity of superlattice reflection (113) decreases, and for $x = 0.12$ composition this peak disappeared.

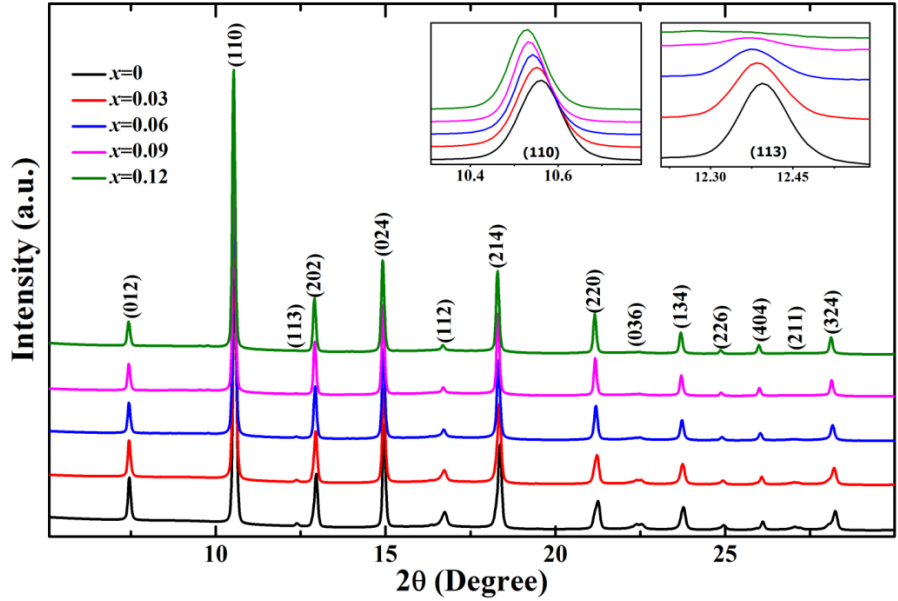

**Figure 1.** *(a) Synchrotron powder XRD pattern of $(Na_{0.5-x}K_xBi_{0.5-x}La_x)TiO_3$, $(0 \leq x \leq 0.12)$, samples showing dominant NBT structure for pure and substituted samples. (Inset) Peak shifting of (110) to lower $2\theta$ with the increasing substitution implying lattice expansion, and (inset) Peak (113) is disappeared with increase in K/La substitution implying that increase in lattice disorder and suppression in octahedral tilting.*

In General crystal structure of NBT is described by rhombohedral unit cell (with *R3c* space group) by cation displacement along [111] with anti-phase tilting of (TiO$_6$) octahedral, whereas, Megwan and Darlingtone chose hexagonal axis ($a_H$ and $c_H$) for refinement of rhombohedral *R3c* unit cell with the lattice transformation as shown in table 1. Where, 't' and 's' are defined as the polar cationic displacement, 'd' as the octahedral distortion and 'e' as rotation of an octahedron face about the triad axis [31].

**Table 1**. Fractional coordinates for the hexagonal setting of rhombohedral (*R3c* space group) perovskite.

| Atom | Wyckoff site | *x* | *y* | *z* |
|---|---|---|---|---|
| Na/K/Bi/La | 6a | 0 | 0 | 0.25+s |
| Ti | 6a | 0 | 0 | t |
| O | 18b | 1/6-2e-2d | 1/3-4d | 1/12 |

In order to find out the structural variation with the substitution, we have carried out Rietveld refinement of SRPXRD data using Fullprof software for all the composition [32], and results are presented in Fig. 2. Voigt axial divergence asymmetry function was used to model the Bragg peaks. The background was estimated by linear interpolation between selected background points. Scale parameter, lattice parameters (*a, b, c*), half-width parameters (*U, V, W*), position parameters of various lattice sites ($x, y, z$) (Table 1) (with z position of O taken fix at 1/12), and thermal parameter $B_{iso}$ (isotropic Debye-Waller factor) were refined one by one. A goodness of refinement is obtained after refining preferred orientation and asymmetric parameters. Acceptable R-factor $R_{wp}$, and $R_{exp}$ were obtained for all compositions. Rietveld refinement of samples with $x = 0$, 0.06 and 0.12 compositions along with R-factors of fit are shown in Fig. 2(a-c). From the Rietveld refinement, it is observed that the lattice parameters '$a$' increases while '$c$' decreases with increase in substitution (as shown in Fig. 2e). However, as shown in Fig. 2e the average unit cell volume increases with increase in substitution. The observed change in lattice parameters and overall volume may be due to the relative increase of effective ionic radii at A-site with increase in substitution, which is a clear indication of successful substitution at A-site of $K^+/La^{3+}$ on the place of atoms $Na^+/Bi^{3+}$ in NBT parent compound. Average crystal radii ($r_A$) of A-site cations in NKBLT-$x$ compositions were calculated using the following equations.

$$r_A = \frac{1}{2} * [(1-x)(r_{Na^+} + r_{Bi^{3+}}) + x(r_{K^+} + r_{La^{3+}})] \tag{1}$$

Where, $r_{Na^+}, r_{K^+}, r_{Bi^{3+}}$, and $r_{La^{3+}}$ crystal radii of $Na^+, K^+, B^{3+}$, and $La^{3+}$, respectively. To estimate the relative distortion in the unit cell with substitution, '$c/a$' ratio was calculated. The '$c/a$' ratio is found to decreases from 2.47 to 2.44 for $x = 0$ and 0.12 composition respectively; it implies that distortion is reduced in the unit cell with the increase in substitution. Distortion is related to the strength of chemical bonding between the constituent elements. Hybridization happens between $Ti(3d)$ and $O(2p)$ electrons. However, a stronger hybridization takes place between lone pair electrons of $Bi(6s^2)$ with O(2p) electrons. This hybridization is responsible for ferroelectricity in NBT structure [16-20]. Both $La$ and $K$ do not have such lone pairs. As a result, the distortive factor is reduced. The estimated thermal parameter $B_{iso}$ of A-site is found to increases from 2.18(3) Å$^2$ to 3.55(5) Å$^2$ for $x = 0$ and 0.12 composition respectively (as shown in Fig. 2f). Which imply that the configurational static disorder in A-site is increasing with the substitution of $K^+/La^{3+}$ at the A-site [33].

While the tilt angle of octahedral 'ω' is given by tanω = $4*(3)^{1/2}$e [34, 35]. Anti-phase ($a^-a^-a^-$) octahedral tilting angle 'ω' decreased from 8.47° to 5.89° for $x = 0$ and $x = 0.12$ composition respectively Fig. 2f (right). Moreover, the $Ti-O$ bond lengths are changes from 2.028 Å and 1.877 Å, to 1.971 Å and 1.938 Å for $x = 0$ and 0.12 composition respectively, it implies that distortion reduced with increase in substitution. In order to visualize changes in atomic positions and orientation of octahedral in a unit cell, the geometric structure is drawn using refined CIF file (Crystallographic Information File) for all composition by VESTA software [36]. As a representative for $x = 0$ and 0.12 structure has been seen from ($x$ $y$ $z$). In Fig 3a and b off-centering (distance between to cross dotted line center of A-site atom) of A-site atoms are shown, for $x = 0$ and 0.12 composition. It is observed that off-centering of A-site cation was more in parent composition comparatively to the $x = 0.12$ composition. Off-centering of A-site cation was played very important role in origin of ferroelectricity in pervoskite materials[37]. In fig. 3c and 3d have been shown octahedra rotation of $x = 0$ and 0.12 composition respectively, from these figures, it can be perceived that the octahedral tilt along the vertical direction in the figure reduced substantially. Figures 3c and 3d are drawn in the *ab*-plain seen from *c* axis [001], in this figure two octahedra's are seen one over the other, shows that the in-plain (*ab*-plain) rotation over the *c* axis reduced as the concentration of *x* increases (highlighted in the black circle).

As observed from SRPXRD, the intensity of Bragg reflection (113) decreases with increase in *x*, and for $x = 0.12$ composition, this peak disappeared. It has been reported that Bragg reflection (113) is related to octahedral tilting and long-range order [25, 34]. In our system, as reported earlier, the tilt angle $\omega$ changes from 8.47° to 5.89° for $x = 0$ to 0.12 composition respectively. All thought, there exist small tilt angle in $x = 0.12$ sample compared to the parent compound, the (113) reflection is disappeared; this may be due to the observed increased disorder at A-site in the system.

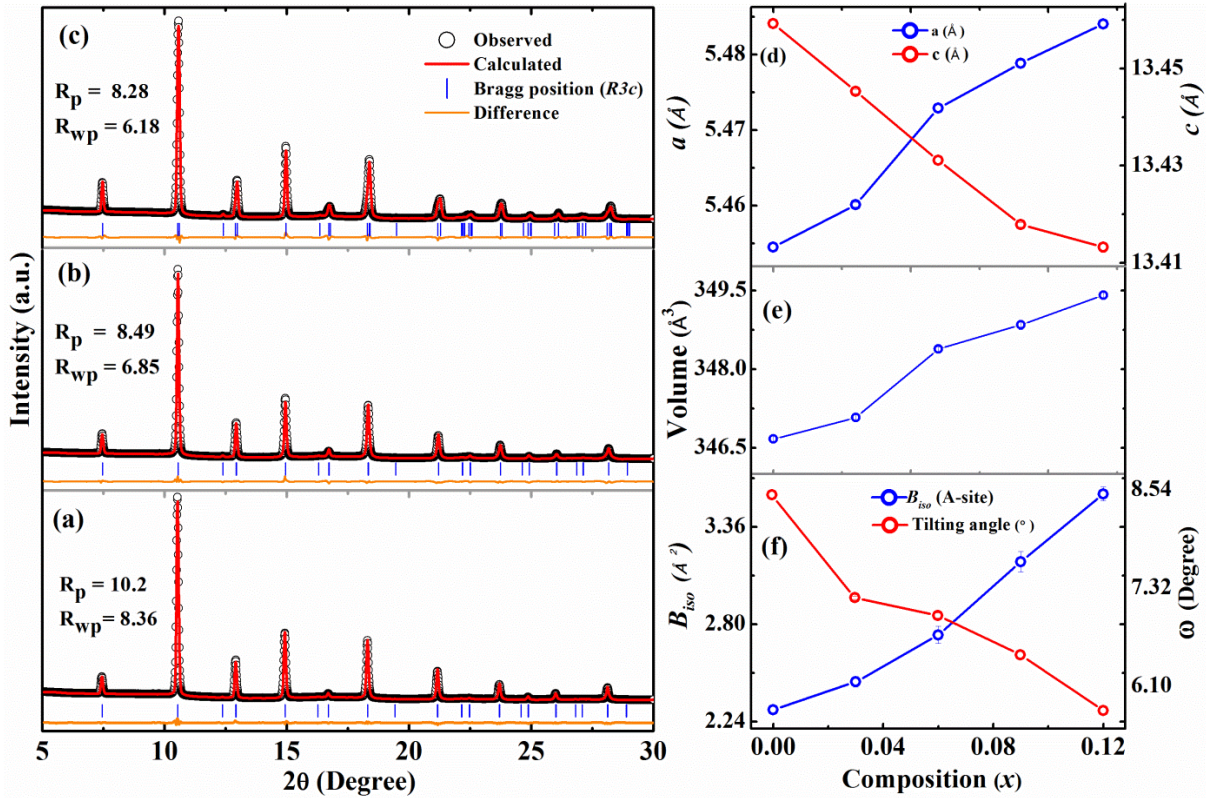

**Figure 2.** *Rietveld refinement of SRPXRD data of $(Na_{0.5-x}K_xBi_{0.5-x}La_x)TiO_3$, ($0 \leq x \leq 0.12$) with goodness of fitting after final cycle with parameters where, (a) $x = 0$, (b) $x = 0.06$ and (c) $x = 0.12$ compositions, (e) Lattice parameter of unit cell 'a' increases while 'c' decreases with substitution, (c) Increase in unit cell volume with substitution, and (d) $B_{iso}$ (isotropic Debye-Waller factors) increases with composition may be lattice system more disordered (right), octahedral tilting angle decreases with composition (left),*

**Table 2.** Atomic positions, and lattice parameters of *NKBLT-x* with (0 ≤ *x* ≤ 0.12) ceramics refinement with *R3c* space group, obtained after final cycle of refinement.

| Atoms | Coordinates | 0 | 0.03 | 0.06 | 0.09 | 0.12 |
|---|---|---|---|---|---|---|
| Na/Bi/K/La | x | 0 | 0 | 0 | 0 | 0 |
| | y | 0 | 0 | 0 | 0 | 0 |
| | z | 0.2730 | 0.2689 | 0.2570 | 0.2562 | 0.2514 |
| Ti | x | 0 | 0 | 0 | 0 | 0 |
| | y | 0 | 0 | 0 | 0 | 0 |
| | z | 0.009 | 0.0087 | 0.006 | 0.0034 | 0.0029 |
| O | x | 0.1265 | 0.1352 | 0.1387 | 0.1431 | 0.1480 |
| | y | 0.3392 | 0.3434 | 0.3481 | 0.3519 | 0.3546 |
| | z | 0.08333 | 0.08333 | 0.08333 | 0.08333 | 0.08333 |

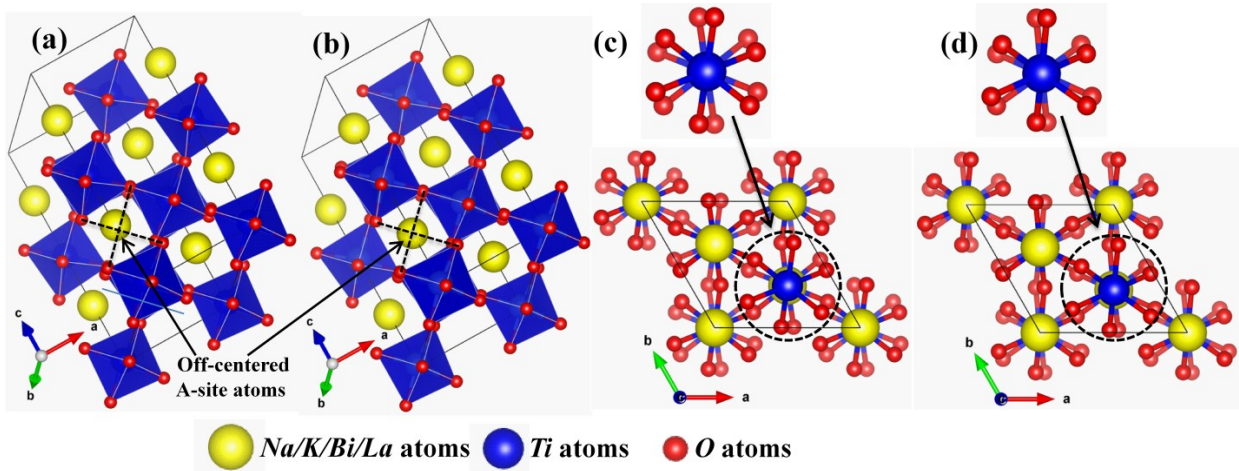

*Figure 3.* Unit cell structure with rhombohedral (R3c space group) of perovskite $(Na_{0.5-x}K_xBi_{0.5-x}La_x)TiO_3$, (0 ≤ *x* ≤ 0.12) *from different perspective, where (a,c) x = 0 composition (b,d) x = 0.12 composition octahedral tilting from projection from (xyz) and (00x) plain respectively, with composition octahedral tilting suppressed.*

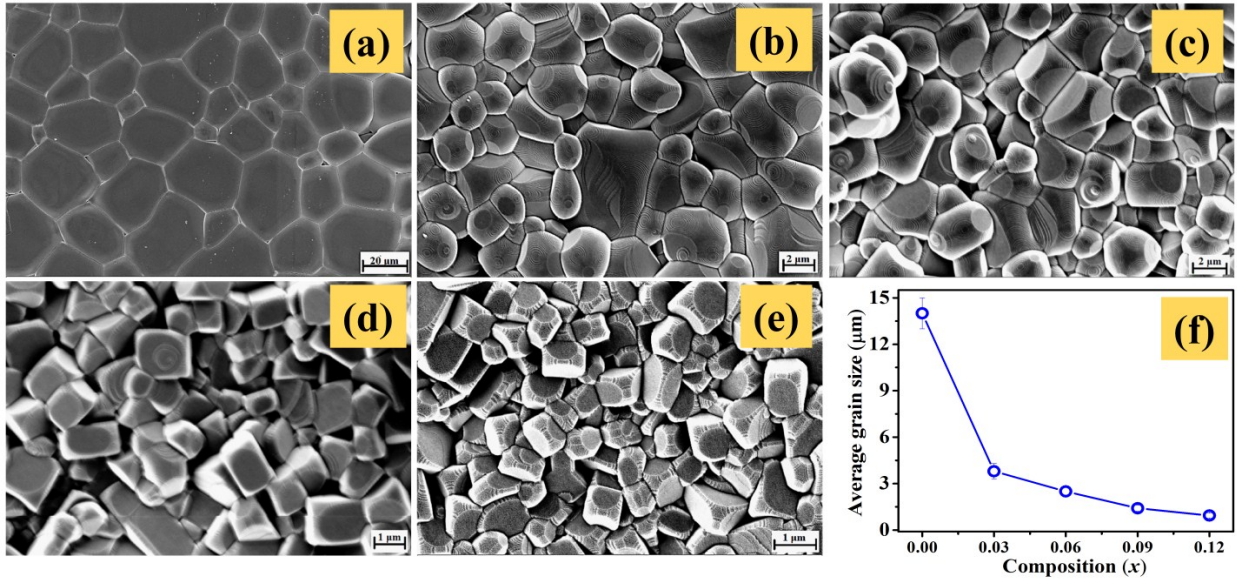

**Figure 4:** *Surface morphology (Na$_{0.5-x}$K$_x$Bi$_{0.5-x}$La$_x$)TiO$_3$, (0 ≤ x ≤ 0.12) compositions of sintered pellets showing reducing grain size of particles where (a) x = 0, (b) x = 0.03, and (c) x = 0.06, (d) x = 0.09, and (e) x = 0.12; (f) Average grain size variation with compositions.*

Surface morphology of the *NKBLT-x* sintered pellets, at 1150 °C for 3 h, was examined using FESEM [Fig. 4(a-e)]. It is clearly visible that, grains are close-packed for all composition. Average grain size was estimated using Image J software and was found to decrease form 14.01±0.99μm to 0.94±0.15μm for $x$ = 0 to 0.12 respectively as shown in Fig.4f. It is well known that rare-earth elements, due to their low diffusivity, are inhibitors of grain growth. Substitution of $K^+$ in the place of $Na^+$ reduces the agglomeration resulted in substantial reduction in particle size. This indicates substituted ions of the large radius are more difficult to diffuse to facilitate grain growth and more efficient to suppress the grain growth than those of small radius [18, 28, 38]. Theoretical density ($\rho_t$) was calculated from the refinement of XRD data, and experimental density ($\rho_e$) was calculated from Archimedes' methods. Relative density ($\rho_R = \rho_e / \rho_t$), continually increasing ( $x(0) \sim 91.5\%$, $x(0.03) \sim 92.4\%$, $x(0.06) \sim 93.7\%$, $x(0.09) \sim 94.9\%$, and $x(0.12) \sim 95.6\%$). Which shows that the suppressed grain growth helps in better densification [39].

EDS was performed on various regions of all the samples. A representative NKBLT-0.12 sample, as shown in Fig. 5(b-g), demonstrates the presence and homogenous distribution of *Na, K, Bi, La, Ti* and *O* without segregation over a selected area (shown in secondary electron micrograph). An integrated area spectrum and estimation of atomic and the weight percentage of constituent elements reveals a close match with stoichiometric composition taken for sample preparation [Fig. 5h].

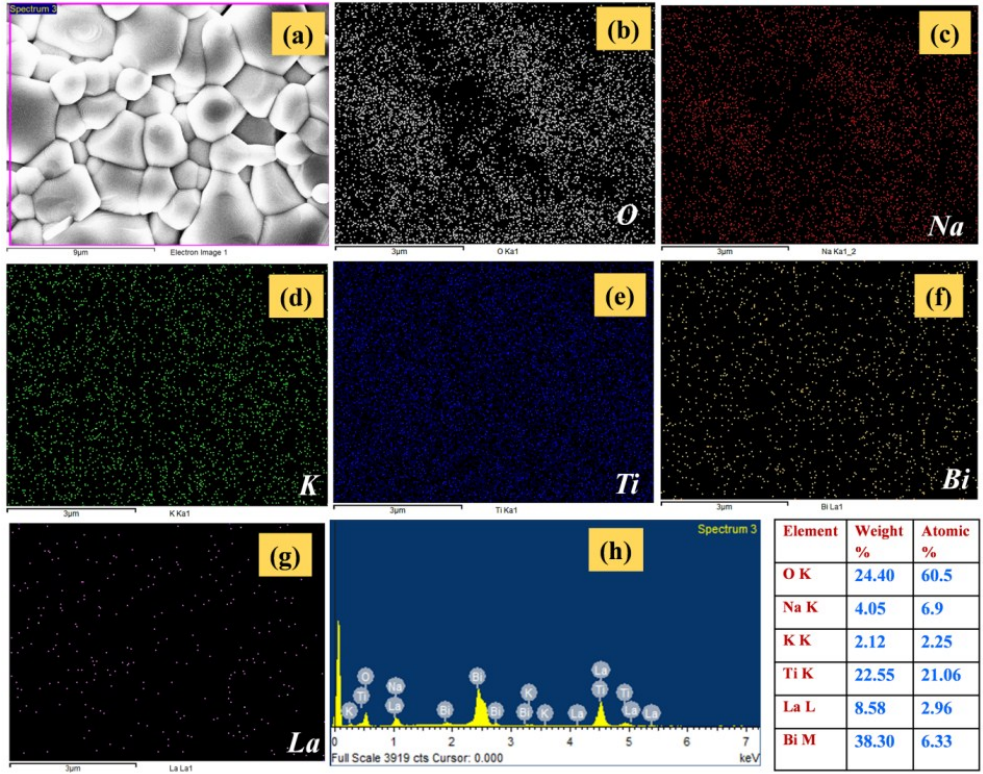

**Figure 5.** *Elemental analysis of $(Na_{0.5-x}K_xBi_{0.5-x}La_x)TiO_3$ with $x$ = 0.12 composition where (a) FESEM image, Elemental mapping of (b) O, (c) Na, (d) K, (e) Ti, (f) Bi, (g) La showing uniform compositional homogeneity. (h) Area EDS data showing all contributing ions also atomic and weight percentage of constituent elements (Table).*

Room temperature frequency-dependent relative permittivity ($\varepsilon_r$) and dielectric loss (tan$\delta$) of *NKBLT-x* ($0 \leq x \leq 0.12$) samples were studied some fixed frequencies in 100Hz to 1MHz range. Both $\varepsilon_r$ and *tan$\delta$* decreases with increase in frequency for all samples but increases with increasing substitution [Fig. 6(a-b)]. Polarization in the dielectric material is the sum of the contribution of electronic, ionic, orientation and space polarization, and it is strongly dependent on frequency. At lower frequency, all polarization contributes easily, but as the frequency is increased different polarization will filter out, this is the possible reason behind to decrease $\varepsilon_r$ with frequency. With increase in composition the dielectric constant ($\varepsilon_r$) increases from ~ 250 to ~1095, and dielectric loss (tan$\delta$) slightly increases 0.024 to 0.065 for *x* = 0 and 0.12 composition respectively at 500 kHz frequency. Increasing $\varepsilon_r$ with increasing substitution may be related to decreasing of the phase transition temperature with the increase in substitution (as shown in Fig. 7). Also, some contribution electronic polarizability ($\alpha_D$) of *La* ($\alpha_D$ ~ 170 – 214) and *K* ($\alpha_D$ ~ 289 – 294) is higher than *Bi* ($\alpha_D$ ~ 50-60) and *Na* ($\alpha_D$ ~ 161 – 163) respectively.

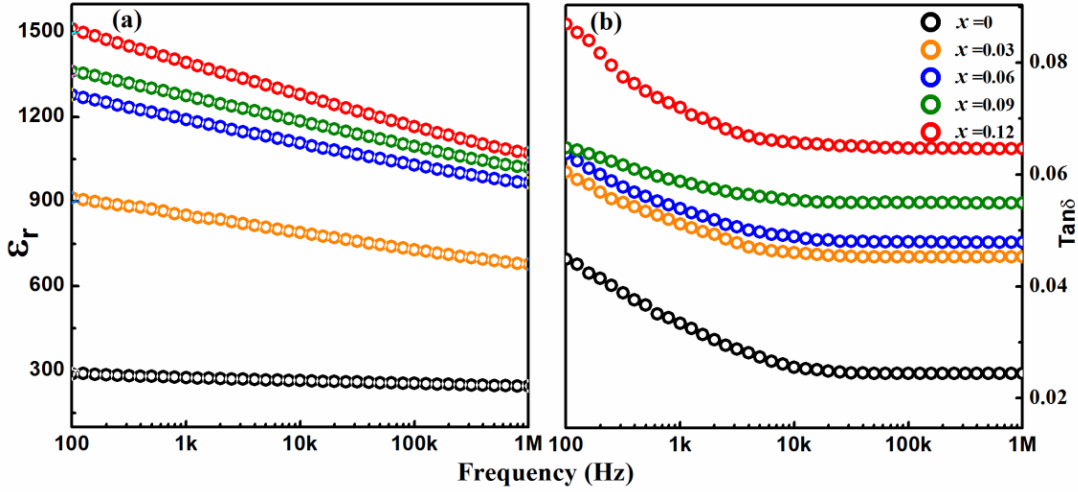

***Figure 6.*** *(a) Frequency-dependence of dielectric relative permittivity ($\varepsilon_r$) of $(Na_{0.5-x}K_xBi_{0.5-x}La_x)TiO_3$, ($0 \leq x \leq 0.12$) at room temperature showing increase of $\varepsilon_r$ with substitution; (b) $\tan\delta$ of the samples increase with the substitution but reduces with frequency.*

Phase transition temperature was investigated using dielectric properties as a probe. The temperature-dependent $\varepsilon_r$ and $\tan\delta$ of *NKBLT-x* samples were measured in the temperature range 50 °C to 450 °C for some fixed frequencies 100Hz to1MHz, results of which are as shown in Fig. 7. Temperature-dependent $\varepsilon_r$ shows that it increases up to a certain temperature and exhibits broad dielectric maxima around $T_m$. After that, $\varepsilon_r$ reduces with increase in temperature above $T_m$. Generally, two broad anomalies observed in NBT, which are $T_d$ and $T_m$. Temperature $T_d$ referred to depolarization temperature and which is corresponding to FE to AFE phase transition temperature. Temperature $T_m$ referred as the maximum dielectric constant temperature which corresponds to AFE to PE phase transition [Fig. 7(a-e)] [40, 41]. In $K^+/La^{3+}$ substituted samples also two diffuse types of anomalies were observed in the temperature-dependent dielectric study, diffuseness of these anomalies increases with increase in substitution. These two phase transitions $T_d$, and $T_m$ move towards lower temperatures with increasing substitution (as shown in Table. 3) from ~ 197 °C and 302 °C to 74 °C and 206 °C, for $x$ = 0 and 0.12 composition respectively. Decreasing $T_d$ and $T_m$ with increasing substitution may be related to the reduction of distortion in lattice system. Distortion is related to the strength of chemical bonding between the constituent elements, the origin of distortion is the lone pair electrons of $Bi(6s^2)$, but *La* and *K* do not have such lone pairs. As a result, the distortive factor is reduced. From XRD data refinement, estimated 'c/a' ratio is also showing that distortion reduces with substitution. As a result, the distortive factor is reduced thereby reducing $T_m$. A comparison of phase transition temperature at 500 kHz for all the samples is shown in Fig.7g.

Diffuse phase transition exhibit a broad dielectric anomaly instead of a sharp dielectric anomaly at the Curie point, like normal ferroelectric materials $BaTiO_3$, $PbTiO_3$, etc. The phase transition characteristics of such materials are known to diverge from the characteristic Curie-

Weiss behavior and phase transition behavior explained by the modified Curie-Weiss law [42-44] given below:

$$\frac{1}{\varepsilon'} - \frac{1}{\varepsilon'_m} = C^{-1}(T - T_m)^\gamma \qquad (2)$$

Where *C* is Curie-Weiss constant and γ gives the degree of diffuseness. For sharp phase transition behavior γ = 1 and for ideal diffuse phase transition γ = 2. The degree of diffuseness was calculated by the least square linear fitting of $ln(\frac{1}{\varepsilon'}-\frac{1}{\varepsilon'_m})$ versus *ln* (*T-T$_m$*) curves at a frequency of 500 kHz for all the *(Na$_{0.5-x}$K$_x$Bi$_{0.5-x}$La$_x$)TiO$_3$* ceramic samples. Value of degree of diffuseness is increasing with substitution from 1.61±0.02 to 1.97±0.02 for *x*=0 and 0.12 composition (Table 3). Diffusions (γ) increased with composition *x*, possible reason behind this may be multiple relaxation processes arising from A-site lattice disorder, which have different transition temperatures. Thus, the enhanced A-site lattice disorder in the samples with K$^+$/La$^{3+}$ substitution as observed in SRPXRD (increase in thermal parameter *B$_{iso}$*) is the reason behind the diffused dielectric phenomenon in the samples [24]. However, as a result, these materials show a region of the steady dielectric constant with temperature. Region for stable dielectric constant, ΔT, is generally given by ε$_{mid}$±15% or better, where ε$_{mid}$ is the value of stable dielectric constant. Temperature limits *T$_{Low}$* and *T$_{High}$* are defined as lower and higher temperatures as (*T$_{Low}$*<*T$_m$* <*T$_{High}$*), and T$_m$ is the temperature corresponding to maximum dielectric constant ε$_m$ [45]. A temperature region, *ΔT*, has been highlighted in yellow for *ε$_{mid}$*±10%. *ΔT* increases with increasing substitution and thermal stability of *ε$_r$* for *x*(0) ~280 °C to 380 °C *x*(0.12) ~ 80 °C to 300 °C. This property makes these materials important to be utilized as high-temperature stable capacitor applications.

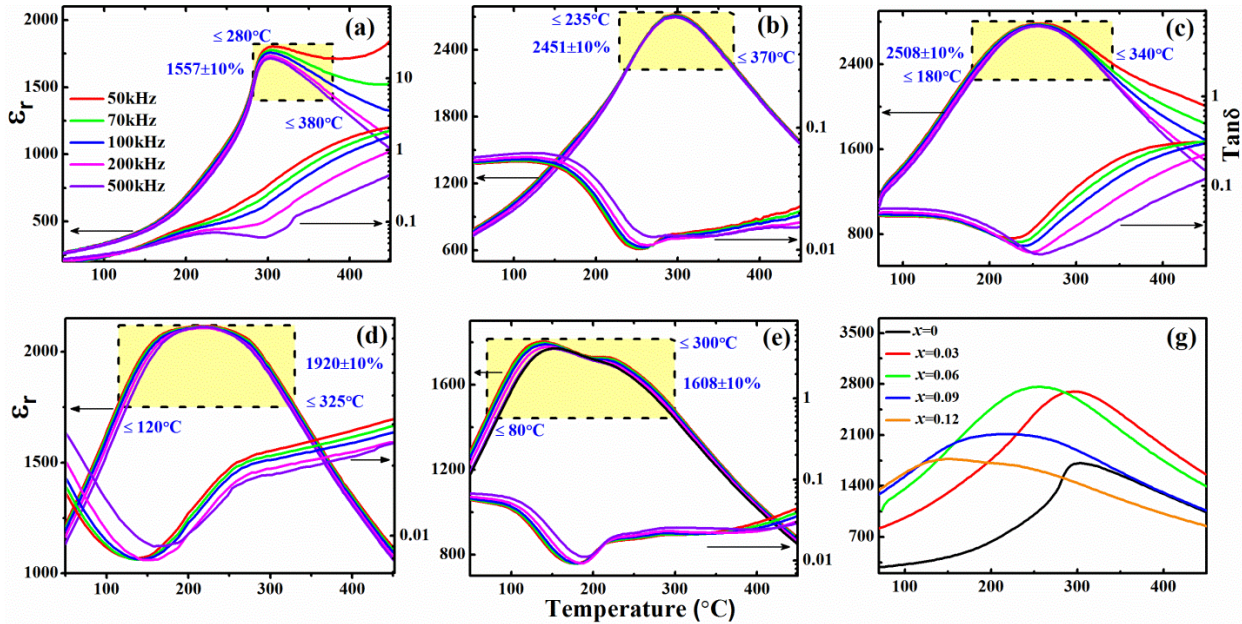

*Figure 7.* Temperature-dependent dielectric measurement of $(Na_{0.5-x}K_xBi_{0.5-x}La_x)TiO_3$, $(0 \leq x \leq 0.12)$ samples for different frequencies; thermally stable region is highlighted (yellow): (a) $x = 0$, (b) $x = 0.03$, (c) $x = 0.06$, (d) $x = 0.09$, and (e) $x = 0.12$, (g) Variation in phase transitions for different substitution at 500kHz.

*Table 3.* Room temperature, high temperature dielectric and ferroelectric properties of $(Na_{0.5-x}K_xBi_{0.5-x}La_x)TiO_3$, $(0 \leq x \leq 0.12)$ compositions

| x | Room Temperature at 500kHz | | $T_d$ (°C) | $T_m$ (°C) | High Temperature at 500kHz | | | | |
|---|---|---|---|---|---|---|---|---|---|
| | $\varepsilon_r$ | Tan$\delta$ | | | $\varepsilon_m$ | Tan$\delta$ | $\gamma$ | $\varepsilon_{mid}\pm10\%$ | $\Delta T$ (°C) |
| 0 | 250 | 0.024 | 197 | 302 | 1015 | 0.063 | 1.61±0.02 | 1557 | 280-380 |
| 0.03 | 690 | 0.045 | 173 | 296 | 2696 | 0.013 | 1.74±0.01 | 2451 | 235-370 |
| 0.06 | 980 | 0.047 | 151 | 253 | 2757 | 0.018 | 1.85±0.03 | 2508 | 180-340 |
| 0.09 | 1037 | 0.054 | 113 | 218 | 2112 | 0.011 | 1.94±0.03 | 1920 | 120-325 |
| 0.12 | 1095 | 0.065 | 74 | 206 | 1769 | 0.021 | 1.97±0.02 | 1608 | 80-300 |

* $\Delta T$ is temperature range for thermally stable dielectric constant $\varepsilon_{mid}\pm10\%$

The dynamic hysteresis with the applied electric field in ferroelectric materials is intrinsically related to the dynamics of the polarization reversal under a cycle of a time-varying electric field, where the hysteretic polarization reversal behaviors are usually induced by domain wall motion and domain switching. Polarization, $P$, was measured as a function of applied *ac* electric field $E$ (maximum electric field for all samples before breakdown), at a frequency of 1Hz at room temperature as shown in Fig. 8(a-e). It is observed from P-E loop measurement that increases in A-site substitution of $K^+/La^{3+}$, significantly effects the coercive field ($E_c$), remnant polarization ($P_r$) and maximum polarization ($P_{Max}$) given in Table 3. Coercive field ($E_c$) monotonously decreases with increasing substitution; it follows the same trend, the grain size with substitution. This reduction in $E_c$ may be related to the reduction of grain size with substitution. Reduced coercive field with substitution may be due to decreased domain size and increased domain switching [46, 47]. In ferroelectric materials, softening of $E_c$ is generally related to the variation in their domain size. Correlation between domain size ($d$) and grain size ($t$) is given by the following equation: [48, 49]

$$d = \left[\left(\frac{\sigma}{\varepsilon^* P_0^2}\right) \cdot t\right] \qquad (3)$$

Where, $\sigma$, $\varepsilon^*$ and $P_0$ are the energy density of the domain wall, effective dielectric constant, and spontaneous polarization, respectively. It may be decrement in grain size is the possible reason behind reduction in $E_c$. Pure NBT has high coercive field and leakage current. Therefore difficult to pole this sample, by $K^+/La^{3+}$ substitution $E_c$ was reduced. Thus easy to pole these samples, it is an advantage of these materials to use in electronics device. However, remnant polarization ($P_r$) first increases till $x = 0.06$ composition, after that it decreases.

P-E loops become slimmer with substitution, it is related to energy storage properties of ferroelectric materials. Recoverable energy density ($W_S$ ~green area in Fig.8e), released during discharging process was calculated from P-E loops [50, 51] using:

$$W_S = \int_{P_r}^{P_{Max}} E dP \tag{4}$$

Where, $E$ denotes the applied external electric field, $P_{Max}$ and $P_r$ are the maximum and remnant polarization respectively. Recoverable energy ($W_S$) increases with the increase in substitution. Hysteresis loss ($W_L$ ~ orange area in Fig. 8e), i.e. unrecoverable energy in the discharging process can be evaluated similarly. Total energy supplied in charging process is $W_{Supplied} = W_S + W_L$. Hence, energy storage efficiency (η) can be defined:

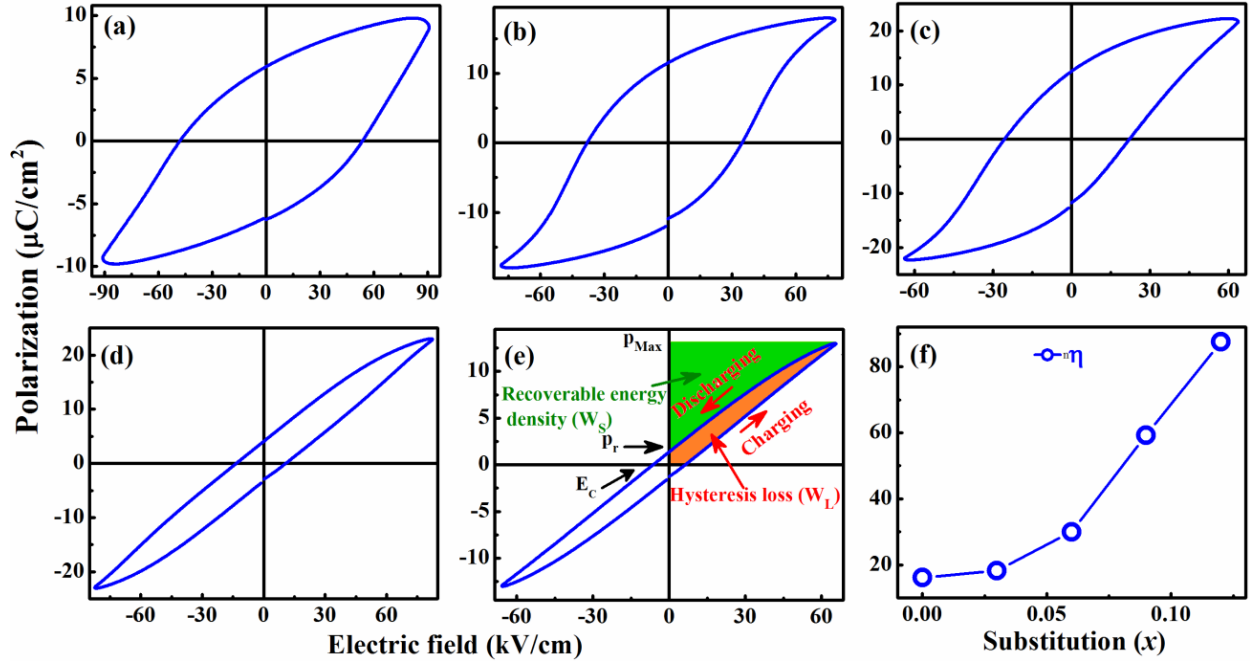

*Figure 8. Ferroelectric properties of $(Na_{0.5-x}K_xBi_{0.5-x}La_x)TiO_3$, $(0 \leq x \leq 0.12)$ samples, where (a) x = 0, (b) x = 0.03, (c) x = 0.06, (d) x = 0.09, (e) x = 0.12 is revealed from polarization verses electric field loops. Figure 8e is marked to indicate different parameters used to calculate energy density, and (f) Energy storage efficiency versus composition.*

$$\text{Energy storage efficiency (η)} = \frac{W_S}{W_{Supplied}} * 100 \tag{5}$$

Energy storage efficiency, η, for NKBLT-x samples [Fig. 8f] increases exponentially with increasing substitution. It is noteworthy to mention that η rises from 17% in NBT to 87 % in $x$ = 0.12 sample. Also, lower composition samples can be used in energy conversion devices etc.

**Table 4**. Ferroelectric properties at room temperature of $(Na_{0.5-x}K_xBi_{0.5-x}La_x)TiO_3$ ($0 \leq x \leq 0.12$) samples.

| $X$ | 2E (kV/cm) | 2$P_r$ (μC/cm$^2$) | Stored energy (J/cm$^3$) | $E_{Max}$ (kV/cm) | η (%) |
|---|---|---|---|---|---|
| 0 | 95.9 | 11.84 | 0.124 | 90.8 | 16.15 |
| 0.03 | 76.28 | 23.09 | 0.182 | 78.2 | 18.26 |
| 0.06 | 51.36 | 25.04 | 0.215 | 63.7 | 30 |
| 0.09 | 26.8 | 8.12 | 0.340 | 84.2 | 59.3 |
| 0.12 | 12.68 | 2.8 | 0.358 | 65.7 | 87.54 |

**Conclusions**

In summary, single-phase lead-free ferroelectric $(Na_{0.5-x}K_xBi_{0.5-x}La_x)TiO_3$ ($0 \leq x \leq 0.12$) successfully synthesized by the sol-gel self-combustion method. Structural analysis of synchrotron powder XRD data by Rietveld refinement for all the samples reveals that all compositions belong to rhombohedral crystal structure with *R3c* space group. There is no structural change in samples with composition. Anti-phase octahedral tilting was reduced from 8.47° to 5.89° for $x$ = 0 and 0.12 composition respectively. Lattice parameters '*a*' are increased, and '*c*' is decreased, while overall volume increased, the clear indication of successfully substituted because of increase in effective A-site substituted radii. Also, lattice distortion is reduced while disorders are increased with increase in substitution. FESEM micrographs analysis observed that all the samples were well dense with reduced in average grain size with substitution. Room temperature frequency dependent study shows that dielectric constant increases with substitution. A temperature-dependent dielectric study revealed FE to AFE, and AFE to PE phase transitions shifted towards lower temperatures and increase in these diffused anomalies with $K^+/La^{3+}$ substitution due to the reduction in distortion with substitution. This provides a stable dielectric constant over a larger temperature range. Thus the materials become more important to be utilized in high-temperature stable capacitor applications due to the enhancement of A-site lattice disorder. Ferroelectricity is observed in all NKBLT-*x* samples. Energy storage efficiency increased with substitution and was highest ~87% for $x$ = 0.12 composition. Result shows that this material has vast scope in high-temperature stable capacitors beyond operating temperature >200°C with high dielectric constant (for $x$ = 0.06 composition with $\varepsilon_{mid} \sim 2508\pm10$ % for the temperature range ~180 °C to 340 °C) and efficient energy storage applications.

**Acknowledgment**

Principle investigator expresses thanks to the Indian Institute of Technology Indore, India for funding the research and using Sophisticated Instrument Centre (SIC). Dr. Sajal Biring acknowledges financial support from the Ministry of Science and Technology, Taiwan (MOST 105-2218-E-131-003 and 106-2221-E-131-027). Dr. Sunil Kumar sincerely thanks SERB for Early Career Research award (ECR/2015/0561).